\documentclass[fleqn,10pt]{wlscirep}
\usepackage[utf8]{inputenc}
\usepackage[T1]{fontenc}
\usepackage{cleveref}  
\usepackage{listings}
\usepackage{orcidlink}  
\usepackage{booktabs}
\usepackage{amsmath}
\usepackage{array}
\usepackage{amssymb}  

\title{{\it infomeasure}: A Comprehensive Python Package for Information Theory Measures and Estimators}

\author[1,*]{Carlson Moses B\"uth\orcidlink{0000-0003-2298-8438}}
\author[1]{Kishor Acharya\orcidlink{0000-0003-3542-6119}}
\author[1]{Massimiliano Zanin\orcidlink{0000-0002-5839-0393}}
\affil[1]{Institute for Cross-Disciplinary Physics and Complex Systems (IFISC), CSIC-UIB, 07122, Palma de Mallorca, Spain}

\affil[*]{carlson@cbueth.de}

\keywords{Information theory, entropy, mutual information, transfer entropy, software package}

\begin{abstract}
Information theory, i.e. the mathematical analysis of information and of its processing, has become a tenet of modern science; yet, its use in real-world studies is usually hindered by its computational complexity, the lack of coherent software frameworks, and, as a consequence, low reproducibility. We here introduce \textit{infomeasure}, an open-source Python package designed to provide robust tools for calculating a wide variety of information-theoretic measures, including entropies, mutual information, transfer entropy and divergences. It is designed for both discrete and continuous variables; implements state-of-the-art estimation techniques; and allows the calculation of local measure values, $p$-values and $t$-scores. By unifying these approaches under one consistent framework, \textit{infomeasure} aims to mitigate common pitfalls, ensure reproducibility, and simplify the practical implementation of information-theoretic analyses. In this contribution, we explore the motivation and features of \textit{infomeasure}; its validation, using known analytical solutions; and exemplify its utility in a case study involving the analysis of human brain time series.
\end{abstract}

\begin{document}

\flushbottom
\maketitle

\thispagestyle{empty}

\section*{Introduction}

Information theory, i.e. the mathematical study of information transmission and processing \cite{reza1994introduction, ash2012information, stone2015information}, has become a cornerstone of modern data analysis, with applications ranging from the study of molecules \cite{eckschlager1994information, nalewajski2006information} to outer space \cite{wing2019applications}. It has further found natural applicability in the study of complex systems \cite{anderson1972more}, i.e. those systems composed of numerous interacting components; any attempt to characterise their dynamics often involves information theory, as the latter is the language of computation, and this defines their nature \cite{gell1995quark}.
Metrics can then be used to describe interactions and information flows between these elements, and, more generally, internal organisation patterns. While the prototypical example of this may be neuroscience \cite{quian2009extracting, wibral2014directed, tononi2016integrated}, relevant applications can be found in fields as diverse as finance \cite{maasoumi1993compendium}, ecology \cite{margalef1973information}, or machine \cite{mackay2003information} and deep learning \cite{roberts2022principles}.

These examples may give the impression that applying information theory measures to a given problem is a simple process; this is nevertheless not the case, with more widespread applications being hindered by four main barriers. Firstly, the measurement of a given information aspect, e.g. uncertainty or information transmission, relies on the use of estimators. These are functions and algorithms that yield an estimation of the probability distributions from which data are generated, using a limited number of observed samples. Specifically, limited theoretical guidelines are available to select the best estimator for a given real problem; using suboptimal ones may result in biases and wrong results. 
Still, this is made more challenging by the second barrier: while multiple software libraries exist, they usually focus on specific measures and/or estimators. The interested reader will find a list of some toolboxes, along with their key characteristics, in Tab. \ref{tab:other_packages}; while it is not exhaustive, and does not include application-specific libraries (e.g. tailored to neuroscience data), it clearly shows the heterogeneity both in metrics and in update frequencies. This variety makes any comparison (e.g. testing multiple estimators) more challenging. The consequence is that a practitioner may only be able to compare a few estimators at best, unless multiple toolboxes are integrated in the processing pipeline.
Thirdly, the large variety of available software tools is an obstacle towards reproducibility, as they may include variations in specific aspects of the implementation that may affect the obtained results.
Lastly, the practitioner will face the problem of the large computational cost of these measures; unless robust and highly optimised algorithms are available, the application to large-scale data sets is challenging at best.

We here present \textit{infomeasure}, a flexible, efficient and open-source Python software library, which addresses these issues by providing a consistent and user-friendly framework for a variety of information-theoretic measures and estimators.
It enables the calculation of measures like entropy, mutual information (MI), transfer entropy (TE), and divergence metrics.
These leverage multiple estimation techniques, including kernel methods, ordinal estimators, and KL / KSG algorithms, thus allowing the users to select the most appropriate approach for their data type and research question.
In addition to standard measures, the package supports conditional variants, generalised MI for multiple variables, and advanced entropy formulations like R\'enyi and Tsallis entropies.
Finally, owing to a modular internal design, processing steps can be customised, and its components can easily be integrated into existing workflows.

By unifying these methodologies, \textit{infomeasure} ensures consistency, reproducibility, and ease of use. Its modular design and extensive documentation make it suitable for a diverse audience, including neuroscientists, machine learning practitioners, and physicists. Most importantly, it has been designed with users with different technical experience in mind; it supports both high-level abstraction and low-level manual configuration, i.e. from simple one-liner executions to detailed component combinations. Researchers can thus focus on their scientific questions, without being burdened by implementation details (unless they want to).
In what follows we describe the main features of \textit{infomeasure} and provide the reader with some basic usage examples. We further present validation experiments using synthetic data for which analytical solutions are known; and illustrate the usefulness of the package in a case study involving the analysis of brain electroencephalographic (EEG) time series.

\begin{table}[!tb]
    \centering
    \caption{Overview of existing information theory toolboxes. This list is non-exhaustive. The last column reports the date of the last human commit on the main branch, as of July 2025. \label{tab:other_packages}}
    \begin{tabular}{p{2.7cm} p{5cm} c p{3.8cm} c}
        \toprule
    Name & Link & Language & Variables / methods & Last update \\
        \midrule

	dit\cite{jamesDitPythonPackage2018} & \url{https://dit.readthedocs.io/en/latest/index.html} & Python & Discrete & May 2025 \\
    DiscreteEntropy.jl\cite{kellyDiscreteEntropyjlEntropyEstimation2024} & \url{https://kellino.github.io/DiscreteEntropy.jl/dev/} & Julia & Discrete & May 2025 \\
    IDTxl\cite{wollstadtIDTxlInformationDynamics2019} & \url{https://github.com/pwollstadt/IDTxl} & Python & MI, TE, AIS, PID & April 2025 \\
	infotheory\cite{candadaiInfotheoryPythonPackage2020} & \url{https://mcandadai.com/infotheory/} & Python/C++ & Discrete/continuous, limited estimators. & August 2020 \\
    InfoTheory.jl\cite{feldtRobertfeldtInfoTheoryjl2023} & \url{https://github.com/robertfeldt/InfoTheory.jl} & Julia & Entropy only & January 2016 \\
	JIDT\cite{lizierJIDTInformationTheoreticToolkit2014} & \url{https://jlizier.github.io/jidt/} & Java & Discrete / Continuous & May 2025 \\
    pyentropy\cite{incePythonInformationTheoretic2009} & \url{https://code.google.com/archive/p/pyentropy/} & Python & Entropy and MI & October 2019 \\
    pyEntropy (\textit{pyentrp})\cite{donetsNikdonPyEntropy2025} & \url{https://github.com/nikdon/pyEntropy} & Python & Entropy & July 2025 \\
    PyInform\cite{mooreELIFEASUPyInformV0202019} & \url{https://pypi.org/project/pyinform/} & Python & Discrete & December 2019 \\
	Pyitlib (MIT)\cite{fosterPyitlibLibraryInformationtheoretic} & \url{https://pypi.org/project/pyitlib/} & Python & Discrete & May 2025 \\
	The Transfer Entropy Toolbox (TET)\cite{hansenTransferEntropyToolbox2013} & \url{https://code.google.com/archive/p/transfer-entropy-toolbox/} & Matlab & Binary time series & June 2013 \\
	TRENTOL\cite{lindnerTRENTOOLMatlabOpen2011} & \url{https://trentool.github.io/TRENTOOL3/} & Matlab & Discrete/continuous, only TE, limited estimators & November 2017 \\

        \bottomrule
    \end{tabular}
\end{table}

\section*{The \textit{infomeasure} package}

\textit{infomeasure} is a Python library designed for efficient and accurate computation of information-theoretic measures. It is compatible with Python versions \texttt{3.11} and following, ensuring broad accessibility. The package is developed following best coding practices, adhering to open-source principles, and includes a comprehensive Code of Conduct (\url{https://github.com/cbueth/infomeasure/blob/main/CODE\_OF\_CONDUCT.md}) and Contributing Guidelines (\url{https://github.com/cbueth/infomeasure/blob/main/CONTRIBUTING.md}), fostering a welcoming and collaborative community.

\subsection*{Measures, estimators, and features}

At the core of the \textit{infomeasure} library are a set of information-theoretical measures, which can be calculated both on discrete and continuous data and time series. 

\begin{itemize}
    \item \textit{Entropy}. Amount of uncertainty associated with a random variable $X$, calculated according to the Shannon's approach. 
    \item \textit{R\'enyi and Tsallis Estimations}. R\'enyi \cite{leonenkoClassRenyiInformation2008} and Tsallis \cite{articleTsallis, Tsallis1998, Tsallis1999} entropies are parametric generalisations of Shannon entropy that introduce tunable parameters $\alpha$ and $q$, respectively, which adjust the sensitivity of the entropy measure to different regions of the probability distribution.
    \item \textit{Joint Entropy}. Amount of information needed to describe two random variables  $X$ and $Y$ together, or equivalently, a measure of the total uncertainty in their joint outcomes.
    \item \textit{Cross-Entropy}. Amount of information needed to encode samples from a distribution $P$ using a code optimized for another distribution $Q$.
    \item \textit{Mutual Information} (MI). Information shared between two random variables $X$ and $Y$; alternatively, it can be interpreted as the average reduction in uncertainty about $X$ that results from learning the value of $Y$, and vice versa.
    \item \textit{Conditional Mutual Information} (cMI). MI between two processes $X$ and $Y$, conditioned on another process $Z$. It provides the shared information between $X$ and $Y$, when considering the knowledge of the conditional variable $Z$.
    \item \textit{Transfer Entropy} (TE). The TE from the source process $X$ to the target process $Y$ is the amount of uncertainty reduced in the future values of $Y$ by knowing the past values of $X$, after considering the past values of $Y$ \cite{schreiber2000measuring}. It thus represents the reduction in the uncertainty in the target variable due to another source variable, that is not already explained by the target variable's past. Equivalently, TE can be interpreted as the amount of information that a source process provides about the target process' next state that was not contained in the target's past states.
    \item \textit{Conditional Transfer Entropy} (cTE). The conditional TE corresponds to the amount of uncertainty reduced in the future values of  $Y$ by knowing the past values of $X$, $Z$, and the past values of $Y$ itself.
    \item \textit{Kullback-Leibler Divergence} (KLD). Mathematical measure of the difference, or of the relative entropy, between two probability distributions $P$ and $Q$. In other words, it can be seen as the degree of surprise one encounters by falsely assuming the distribution $P$ instead of the true distribution $Q$ in a model.
    \item \textit{Jensen-Shannon Divergence} (JSD). Measure of the dissimilarity between two probability distributions \cite{lin1991divergence}; also, symmetric version of the KLD.
\end{itemize}

As previously introduced, these measures can only be calculated by knowing the probability distributions underlying the observed data. As these are usually not known, except for special cases and toy models, the solution requires the use of estimators, i.e. functions and algorithms that yield an estimation of the probability distributions from which data are generated. The \textit{infomeasure} library supports a large set of them, whose main families are listed and described below. Additionally, most combinations of estimators and measures are natively supported, see Tab. \ref{tab:measures}. Finally, a short guide for the user new to these approaches is provided at \url{https://infomeasure.readthedocs.io/en/latest/guide/estimator_selection/}.

\begin{itemize}

    \item \textit{Kernel Estimation}. Kernel estimation techniques employ kernel density estimation (KDE) to approximate the probability density functions (pdf) necessary for computing information-theoretic measures \cite{schreiber2000measuring}. At its core, KDE estimates the density at a given point by averaging contributions from all sample points, weighted by their distance, using a kernel function \cite{silverman1986density}. The choice of kernel (e.g., Box, or Gaussian) and of the bandwidth parameter play a crucial role in determining the smoothness and accuracy of the estimated density and, therefore, also of the information-theoretic measures. 

    \item \textit{Kozachenko-Leonenko (KL) / Kraskov-Stoegbauer-Grassberger (KSG) / Metric / kNN Estimation}. Kozachenko-Leonenko (KL) estimator leverages nearest-neighbour distances within the sample data to provide a direct, asymptotically unbiased, and consistent estimator of differential entropy, circumventing the need for explicit density estimation \cite{kozachenko1987sample}. The idea is to approximate the local density around a sample using the volume of the surrounding ball defined by its $k$-th nearest neighbour. However, when estimating mutual information (MI), decomposing this into marginal and joint entropies using KL estimators can introduces biases, due to inconsistent local density scales across spaces of differing dimensionality. The Kraskov-St\"ogbauer-Grassberger (KSG) estimator addresses this issue by computing nearest-neighbour distances in the joint space, then projecting the associated distance scale into marginal subspaces to count neighbours there \cite{miKSG2004}. This procedure ensures coherence between joint and marginal estimates, improving both bias correction and data efficiency. Variants of the KSG framework have also been extended to conditional MI and transfer entropy.
   
    \item \textit{Ordinal / Symbolic / Permutation Estimation}. Symbolic estimation transforms time series or sequential data into a discrete symbolic sequence using a symbolization rule; for instance, the ordinal patterns method encodes each point into a symbol based on the relative ordering of its local neighbourhood \cite{PermutationEntropy2002}. This transformation captures essential structural features of the data while reducing sensitivity to noise. Once the time series is symbolised, the probability distribution over the symbol space is estimated using relative frequencies, which is subsequently used to estimate the information-theoretic measures.

    \item \textit{Bias-corrected estimators}. Set of estimators substituting the plug-in one for the Shannon's entropy, designed to provide a lower bias for small sample sizes. These estimators have been selected based on the review included in Refs. \cite{de2024entropy, contreras2021selecting}.

\end{itemize}

\begin{table}[!tb]
    \centering
    \caption{Information Measures and Estimation Methods. Measures marked with an asterisk (*) can also be computed using Rényi and Tsallis entropy formulations, in addition to the standard Shannon entropy.
    The non-standard cross-entropy notation has been borrowed from Christopher Olah's blog post ``Visual Information Theory'' \cite{olahVisualInformationTheory}. We choose this nomenclature $H_Q(P)$, as the widely used $H(P, Q)$ is ambiguous with joint entropy. For the bias-corrected estimators, measures marked with a dagger ($\dagger$) are only available for a subset of the estimators.\label{tab:measures}}
    \begin{tabular}{@{}l l c c c c c@{}}
        \toprule
        \textbf{Measures} & \textbf{Notation} & \textbf{Discrete} & \textbf{Kernel} & \textbf{Metric / kNN} & \textbf{Ordinal} & \textbf{Bias-corrected}\\ 
        \textbf{Estimator} & & \textbf{Estimator} & \textbf{Estimator} & \textbf{Estimator} & \textbf{Estimator} & \textbf{Estimators}\\ 
        \midrule
        Shannon Entropy & $H(X)$ & \checkmark & \checkmark & \checkmark & \checkmark & \checkmark \\
        R\'enyi \& Tsallis entropies & $H(X)$ & & & \checkmark & & \\
        Joint Entropy* & $H(X,Y)$ & \checkmark & \checkmark & \checkmark & \checkmark & \checkmark \\
        Cross-Entropy* & $H_Q(P)$ & \checkmark & \checkmark & \checkmark & \checkmark & \checkmark$\dagger$ \\
        Mutual Information (MI)* & $I(X;Y)$ & \checkmark & \checkmark & \checkmark & \checkmark & \checkmark \\
        Conditional MI* & $I(X;Y|Z)$ & \checkmark & \checkmark & \checkmark & \checkmark & \checkmark \\
        Transfer Entropy (TE)* & $T_{X \to Y}$ & \checkmark & \checkmark & \checkmark & \checkmark & \checkmark \\
        Conditional TE* & $T_{X \to Y|Z}$ & \checkmark & \checkmark & \checkmark & \checkmark & \checkmark \\
        Kullback-Leibler Divergence (KLD)* & $\operatorname{KLD}(P||Q)$ & \checkmark & \checkmark & \checkmark & \checkmark & \checkmark$\dagger$ \\
        Jensen-Shannon Divergence (JSD) & $\operatorname{JSD}(P||Q)$ & \checkmark & \checkmark &  & \checkmark & \checkmark$\dagger$ \\ 
        \bottomrule
    \end{tabular}
\end{table}

Some additional features of the library ought to be here highlighted. From the viewpoint of usability, the package unifies the slicing mechanism for the TE estimators, ensuring consistency across all approaches and minimising potential points of failure. It can be used to obtain local values for the entropy, MI, and TE; thus allowing to describe the evolution of the information transfer through time \cite{Lizier2014}. It further simplifies hypothesis testing, by yielding $p$-values, $t$-scores and confidence intervals for MI and TE. Finally, the TE implements an Effective TE (eTE) variant, reducing biases from finite sample sizes \cite{articleKantz}.

From an architectural viewpoint, the library is coded using a robust inheritance-based design, which guarantees reusability and maintainability of the codebase. This, combined with a comprehensive test suite and CI/CD pipelines, further ensures reliability and reproducibility. Finally, a special effort has been devoted to complement the package with a comprehensive documentation, collecting and clarifying various names that have been used for equivalent methods (e.g., Kozachenko-Leonenko (KL) / Metric / kNN Entropy, or Ordinal / Symbolic / Permutation Entropy). 

\begin{figure}[!tb]
\centering
\includegraphics[width=\linewidth]{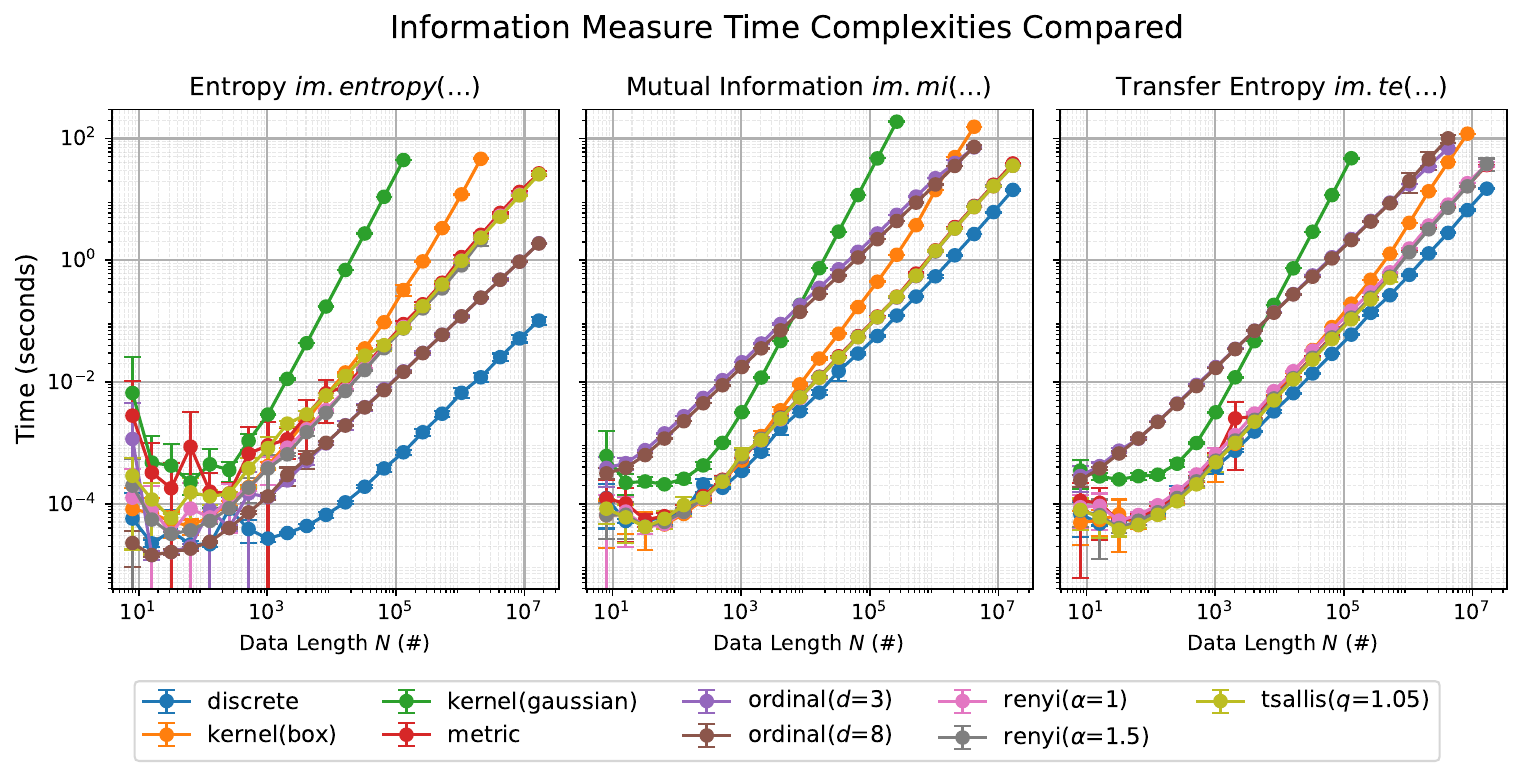}
\caption{Evolution of the computational cost as a function of the time series length. Results have been obtained with CPython 3.13.2 and Clang 18.1.8 (Darwin 24.4.0) on an Apple M4 Pro; only one core used in the computations. Points and whiskers respectively indicate the average and standard deviation over ten independent realisations. Equivalent results for bias-corrected estimators are available at \url{https://infomeasure.readthedocs.io/en/latest/demos/Time_Performance/}.}
\label{fig:comp_cost}
\end{figure}

As a final issue, no software library is truly useful unless it guarantees computational efficiency. Due to the heterogeneity of implementations, a direct comparison with similar libraries is not practical; still, an overview of the computational cost of different measures and estimators is reported in Fig. \ref{fig:comp_cost}. It can be appreciated that, even in the worst scenarios, time series of $10^5$ elements can be analysed in less than a minute in a standard computer.

\subsection*{One-liners and examples}

As previously explained, one of the principles behind the creation of the \textit{infomeasure} package is the simplification of complex calculations through easy-to-use utility functions. In other words, even though an experienced user can delve into the classes and access individual functions, a less experienced one can still perform most computations by calling high-level functions. We are here going to illustrate this point with three basic examples. Firstly, calculating the entropy for a vector of symbols is as simple as calling the \texttt{im.entropy()} function:

\begin{lstlisting}[language=Python, caption={Calculating entropy using the kernel approach, specifying bandwidth and kernel type.}, label={lst:im.entropy.kernel}]
    import infomeasure as im
    data = [1.24, 0.92, 1.87, 1.51, 0.48, 3.60, 1.32]
    entropy = im.entropy(data, approach="kernel", bandwidth=2, kernel="box")
\end{lstlisting}

Moving to a bivariate case, the MI between two (or more) random variables can be computed using \texttt{im.mutual\_information()}. Similarly to the previous case, we have:

\begin{lstlisting}[language=Python, caption={Calculating MI from discrete data.}, label={lst:im.mi.discrete}]
    data_x = [0, 1, 0, 1, 0, 1, 0, 1]
    data_y = [0, 1, 0, 1, 0, 0, 0, 0]
    mi = im.mutual_information(data_x, data_y, approach="discrete")
\end{lstlisting}

Finally, the computation of the TE can be performed though the \texttt{im.transfer\_entropy()} function, as shown below:

\begin{lstlisting}[language=Python, caption={Calculating TE using the metric approach, with noise handling.}, label={lst:im.te.metric}]
    source = [0.0, 0.3, 0.5, 1.2, 0.0, 0.4, 0.2, -0.6, -0.8, -0.4]
    dest = [0.0, 0.8, -0.7, 0.2, 1.2, 1.0, 1.3, 0.7, 0.8, -0.1]
    te = im.transfer_entropy(source, dest, approach="metric", noise_level=0.001)
\end{lstlisting}

In synthesis, these examples demonstrate how \textit{infomeasure} reduces the complexity of using information-theoretic measures. In the simplest scenario, the three functions described above only require the definition of the input data, the approach to use, and eventually the associated parameters. The interested reader will find more specific settings, as well as the meaning of all parameters, in the documentation.

\subsection*{Internal structure overview}

The internal architecture of the \textit{infomeasure} package has been designed to simplify the computation of information-theoretic measures while remaining flexible and extensible. At its core, the package organizes the computation of measures such as entropy, MI, and TE into a set of reusable and modular estimators. These estimators support a variety of approaches, see Tab. \ref{tab:measures}. The package builds on this core functionality to address challenges like selecting the appropriate estimator for a given dataset or integrating multiple measures into a single workflow, as all approaches share the same interface. It ensures that even users with limited technical expertise can compute these measures efficiently through its high-level functional API, as demonstrated in Listings \ref{lst:im.entropy.kernel}, \ref{lst:im.mi.discrete}, and \ref{lst:im.te.metric}. These examples illustrated how users can compute entropy, MI, and TE with just a few lines of code, without needing to understand the underlying implementation details.

The package's flexible design is built around a robust inheritance-based framework. A central `Estimator` base class defines the interfaces for all estimators, ensuring consistency in how data are processed and results are generated. Additional functionality is provided through mixins, such as hypothesis testing (\texttt{PValueMixin}) and effective value computation (\texttt{EffectiveValueMixin}). These mixins enable features like permutation tests and effective transfer entropy to be added seamlessly without duplicating code. To support advanced users, the package also allows direct access to specific estimator classes, providing fine-grained control over parameters and computations.

Another key feature of \textit{infomeasure} is its dynamic import mechanism, implemented in the `functional.py` module. By mapping user-friendly terms such as "kernel" or "ordinal" to their corresponding estimator classes, it minimizes memory usage and ensures that only the necessary components are loaded for a given task. Configuration settings are centralised in the `Config` class, which allows global parameters like the logarithmic base (e.g., bits, nats, hartleys) to be set and adjusted. This ensures consistency across computations while also allowing local overrides for specific use cases. Additionally, logging is handled centrally, making it easier for users to debug and monitor the package's operations.

The package also emphasizes computational efficiency and scalability. The modular design allows new measures or estimation techniques to be integrated easily, ensuring that the library can evolve alongside advancements in the field of information theory. Comprehensive documentation and consistent naming conventions further enhance usability, enabling users to quickly locate and understand the functionality they need. The package is openly accessible via PyPI (\url{https://pypi.org/project/infomeasure/}) and conda-forge (\url{https://anaconda.org/conda-forge/infomeasure}), ensuring broad compatibility and ease of installation.

\section*{Validation}

Beyond a technical verification of all modules and functions using a unit testing approach, the library has been numerically validated using a set of test cases. These involved comparing the values yielded by different measures against the analytical solutions, for random variables following known distributions.

\begin{figure}[ht]
\centering
\includegraphics[width=\linewidth]{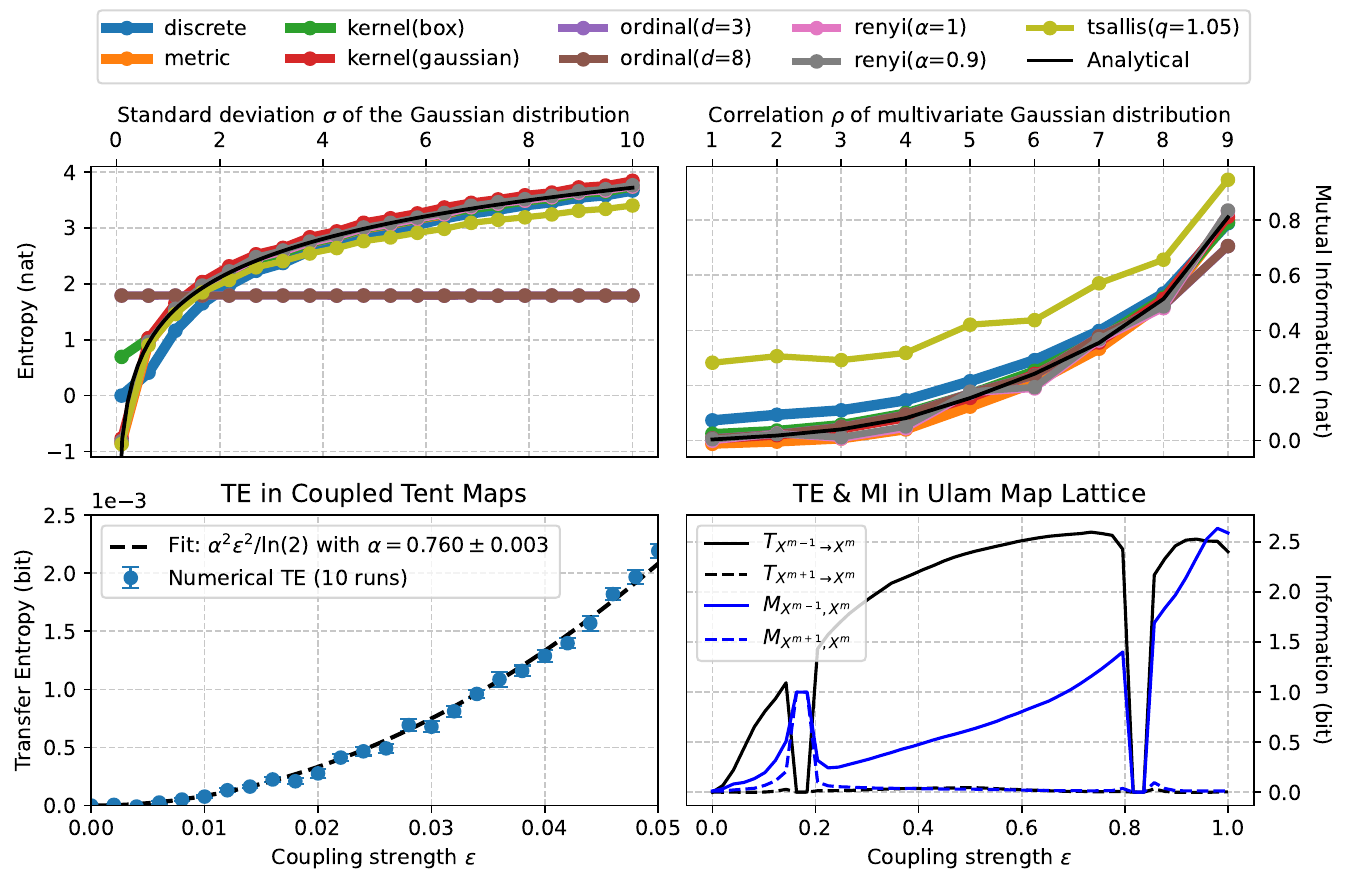}
\caption{Validation of measures and estimators. From left to right, the two top panels report the numerical and analytical evolution of the entropy and the MI, for Gaussian random variables. The two bottom panels report the evolution of the TE as a function of the coupling strength $\epsilon$ for respectively tent map lattices and Ulam map lattices \cite{schreiber2000measuring}. See main text for definitions and details. }
\label{fig:validation}
\end{figure}

We start with the simplest case of a random variable $X$ defined by a normal distribution of zero average and $\sigma$ standard deviation; its maximum likelihood entropy is known to be equal to $H(X) = 1/2 \log( 2 \pi e \sigma^2 )$. The top left panel of Fig. \ref{fig:validation} reports the evolution of such entropy as a function of $\sigma$ (black line), and of the results obtained by different estimators. All methods yield very good approximations of the analytical values, the only exceptions being the ordinal and discrete estimators. In the case of the former, the error comes from the discretisation of values to the nearest integer, which destroys all information for small values of $\sigma$. On the other hand, the ordinal symbolisation process only considers the relative amplitude of values within a pattern, but not their absolute amplitude; in other words, the obtained entropy is independent of the scaling introduced by $\sigma$, hence the constant value. Finally, the box kernel estimator deviates for very small standard deviations, resulting from the kernel bandwidth being too small to capture the information; yet it recovers the analytical curve for larger values.

We next move to a bivariate case, by evaluating the MI between two Gaussian random variables $X$ and $Y$, both with zero mean and unit variance, and with a linear correlation coefficient between them of $\rho$. The theory indicates that the MI between them is given by $I(X; Y) = -1/2 \log( 1 - \rho^2 )$. Again, results obtained by the estimators match the theoretical prediction, see top right panel of Fig. \ref{fig:validation}. In this case the exception is given by Tsallis' MI, here due to the use of $q = 1.05$--note that, for $q = 1$, the Tsallis' MI becomes identical to the Shannon's one.

We finally reproduce the numerical experiments described in T. Schreiber's seminal paper \cite{schreiber2000measuring}, focusing on two canonical systems: the tent map lattice and the Ulam map lattice.
The former case is composed of $100$ coupled tent maps in a 1D lattice topology, the dynamics of the $m$-th element being given by:

\begin{equation}
	x^m = f( \epsilon x^{m-1} + ( 1 - \epsilon ) x^m ),
	\label{eq:lattice}
\end{equation}

with $f$ being the tent map function:

\begin{equation}
	f(x) =
	\begin{cases}
		2 x & \text{if } 0 \le x < 1/2 \\
		2 - 2x & \text{if } 1/2 \le x \le 1.
	\end{cases}
\end{equation}

When each state $x$ is binarised using a threshold of $0.5$, the TE between each site and the next one can be approximated, for small values of the coupling constant $\epsilon$, by:

\begin{equation}
	T_{X^{m-1} \rightarrow X^m} \approx \frac{\alpha^2 \epsilon^2}{\ln 2},
\end{equation}

with $\alpha \approx 0.77$. The bottom left panel of Fig. \ref{fig:validation} reports both the numerical estimations of the TE according to \textit{infomeasure} (blue points, with whiskers representing the standard deviation over $10$ independent realisations), and the corresponding least squares fit. The latter yields a value of $\alpha = 0.760 \pm 0.003$, i.e. very close to the one reported in Ref. \cite{schreiber2000measuring}.

The second considered system is a set Ulam maps, coupled according to the same topology as in Eq. \ref{eq:lattice}, with $f$ now being $f(x) = 2 - x^2$. Four measures are here calculated, i.e. the TE and the MI between pairs of adjacent sites in both directions: $T_{X^{m-1} \rightarrow X^m}$, $T_{X^{m+1} \rightarrow X^m}$, $I(X^{m-1}; X^m)$, and $I(X^{m+1}; X^m)$. Results, reported in the bottom right panel of Fig. \ref{fig:validation}, show that the TE reflects the unidirectional nature of the coupling, while the MI tends to capture static correlations and becomes symmetric in regimes of partial synchronisation.
 
The interested reader can find the code to reproduce these four results, as well as some additional considerations, in the documentation of the package, see \url{https://infomeasure.readthedocs.io/en/0.5.0/demos/gaussian_data/} and \url{https://infomeasure.readthedocs.io/en/0.5.0/demos/Schreiber_Article/}.

\section*{Use case example: analysis of EEG time series}

As a final topic and example of use case, we here describe the task of analysing information transmission in human brain time series. This section is independent from the remainder of the text; the reader not interested in it can safely skip to the conclusions. The underlying idea, at the foundation of the so-called ``network neuroscience'' field \cite{frohlich2016network, bassett2017network}, is that brain activity can be represented by networks; their nodes then map different regions of the brain, and a pair of them are connected by links whenever a statistical dependency is observed between the corresponding dynamics \cite{bullmore2009complex, park2013structural, sporns2013structure}. We here reconstruct functional networks by starting from electroencephalographic (EEG) recordings of 14 patients suffering from schizophrenia, and of the same number of matched control subjects \cite{olejarczyk2017graph}, freely available at \url{http://dx.doi.org/10.18150/repod.0107441}. Data correspond to a resting state, i.e. with the subjects doing no cognitive tasks and with their eyes closed; 19 electrodes were recorded at 250Hz, following the standard positions: Fp1, Fp2, F7, F3, Fz, F4, F8, T3, C3, Cz, C4, T4, T5, P3, Pz, P4, T6, O1, O2. Time series have not been preprocessed and have been used as provided \cite{olejarczyk2017graph}.

\begin{figure}[!tb]
\centering
\includegraphics[width=\linewidth]{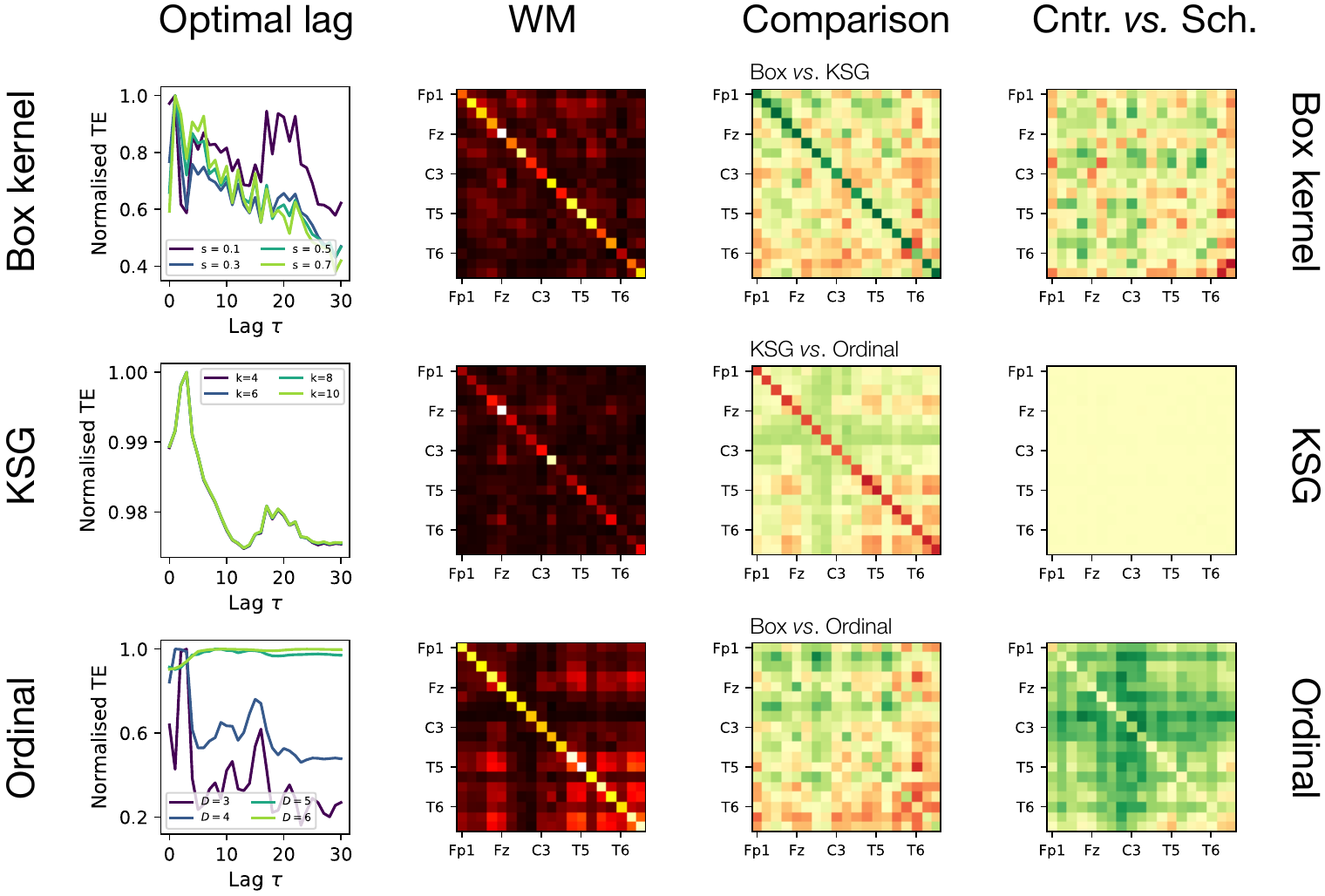}
\caption{Analysis of EEG time series. From left to right, panels correspond to: the evolution of the normalised TE as a function of the lag $\tau$; the obtained TE for each pair of EEG electrodes, with light shades indicating larger TE values; the differences between pairs of estimators, for all possible pairs of EEG electrodes; and the differences between control subjects and patients. From top to bottom, panels correspond to the box kernel, the KSG, and the ordinal estimators. See main text for definitions and details. }
\label{fig:eeg}
\end{figure}

We start by comparing the results obtained by calculating the Transfer Entropy measure between pairs of time series, using three different estimators: the box kernel, the KSG, and the ordinal one. We specifically analyse the average information detected as a function of the lag $\tau$, for different parameters and across $100$ random pairs of time series. Results are reported in the left panels of Fig. \ref{fig:eeg}; in order to simplify comparisons across different parameter values, results have been normalised according to the largest TE detected for each parameter, such that the maximum is always one. Several interesting ideas can already be drawn. Firstly, changing the parameter of each estimator has a varying effect: it strongly modifies the results in the box kernel and ordinal cases, but has no impact in the KSG case. Secondly, in all cases two maxima are observed, one for small $\tau$s, and a second one around $\tau \approx 15$. While the former may be the result of the smoothness of the time series, the latter probably represents a more genuine information transfer. We then choose the parameters yielding the clearest peak: $\tau = 17$ and $s = 0.1$ for the box kernel; $\tau = 17$ and $k = 4$ for KSG; and $\tau = 15$ and $D = 4$ for the ordinal estimator.

When all possible pairwise TEs are calculated, averaged over the $14$ control subjects, the results are those represented in the panels in the second column of Fig. \ref{fig:eeg}. For each pair of EEG electrodes, colours indicate the intensity of the TE between them, from weak (dark shades) to strong (light shades). While some differences can be appreciated, these are more evident in the panels of the third column, where pairs of estimators are compared--see titles on top of each panel, with green (respectively, red) shades indicating a higher (lower) TE for the first estimator. The largest differences can be found for the ordinal estimator, which presents higher TE values between electrodes in the parietal and occipital lobes, and lower values in frontal and temporal lobes.

Panels in the rightmost column finally report the differences that each estimator observes between control subjects and patients. Given the average value observed in each group and for each pair of electrodes $i$ and $j$, the difference has been estimated as $\delta_{i,j} = \log_2 \overline{TE_{i,j}^{cntr}} / \overline{TE_{i,j}^{sch}}$. Positive values of $\delta_{i,j}$ thus indicate that the TE is on average higher in control subjects (represented as green squares); conversely, negative values (red squares) indicate a higher value for patients. Estimators here have completely different behaviours: the box kernel detects differences, but no clear patterns in them; the KSG, minimal differences between the two groups; and the ordinal estimator, a substantially higher TE for the control subjects across all regions, especially marked in the central lobe. Note that this latter result is aligned with previous findings in the literature \cite{harmah2020measuring, jia2022abnormal, wang2024networked}.

In short, this use case illustrates some of the advantages of the \textit{infomeasure} library. On the one hand, it simplifies the computation of information-theoretical measures, in the limit that only one line of code had to be changed to alternate approach and parameters to obtain all results presented in Fig. \ref{fig:eeg}. On the other hand, such flexibility enables comparisons beyond what is usually found in the literature: as an example, we here showed that different estimators observe different patterns discriminating between control subjects and schizophrenic patients. In any case, the reader should note that this is a basic analysis that does not substitute more detailed studies, which would require better data pre-processing and artefact elimination, statistical analysis of the differences, and so forth \cite{harmah2020measuring, jia2022abnormal, wang2024networked}.

\section*{Discussion and conclusions}

We here presented \textit{infomeasure}, a comprehensive Python library for the efficient and accurate computation of information-theoretic measures. Compared to alternative libraries, it presents multiple advantages, including: the availability of numerous estimators, and the possibility of plug them in the calculation of numerous measures (see Tab. \ref{tab:measures}); a modular structure that adapts to the expertise of the user, allowing to get results from one line of code, or by delving deeper into individual modules; the possibility of obtaining local values for the measures; and the support of hypothesis testing, through $p$-values and $t$-scores for MI and TE. While some elements are not automatised and are left to the control of the user (as e.g. the selection of the best parameters), the structure of the library simplifies its integration into larger data analysis pipelines. Additionally, we plan to continuously expand the set of metrics and estimators - the interested reader can find a future development roadmap in the documentation.

These advantages have been illustrated in a use case involving the analysis of EEG time series. One aspect of information-theoretical measures that is often neglected is that they require knowledge of the probability distributions underlying the data; and that, when this is not available, it has to be estimated. As seen in Fig. \ref{fig:eeg}, while different estimators yield a similar global picture, at least in terms of the time scale $\tau$ of the propagation of information, they do not agree in the specific differences between control subjects and patients. By allowing the practitioner to compare a large set of estimators, and by doing that efficiently (both computationally, and in terms of the length of the required code), \textit{infomeasure} allows alleviating this problem. Insofar different estimators leverage and describe different properties of the data, such comparisons could also be used as a novel way of extracting insight on altered brain dynamics.

\bibliography{refs}

\section*{Acknowledgements}

This project has received funding from the European Research Council (ERC) under the European Union's Horizon 2020 research and innovation programme (grant agreement No 851255). This work was partially supported by the Mar\'ia de Maeztu project CEX2021-001164-M funded by the  MICIU/AEI/10.13039/501100011033 and FEDER, EU.

\section*{Author contributions statement}

C.M.B. developed the \textit{infomeasure} package, including design, coding, and validation. K.A. and M.Z. conceptualised the project, guided the implementation, and contributed to validation and testing.
All authors collaborated on the development of demos and applications.
All authors reviewed and approved the manuscript.

\section*{Additional information}

\textbf{Accession codes}: The \textit{infomeasure} package (version 0.5.0) is archived on Zenodo, at \url{https://zenodo.org/records/15792893}.\\
\textbf{Documentation}: The full documentation is available at \url{https://infomeasure.readthedocs.io/}.\\
\textbf{Source Code}: The source code is publicly available on GitHub: \url{https://github.com/cbueth/infomeasure}.  
\textbf{Competing interests}: The authors declare no competing interests.

\section*{Data availability}

The dataset analysed during the current study is available in the RepOD repository, \url{http://dx.doi.org/10.18150/repod.0107441}.

\end{document}